# Spermatogonial Stem Cell and TGF-β involved Regulation of Proliferation and Differentiation

Ping Huang[1], Tingting Wang[2]

[1.] Department of Life Science, Sichuan University, Chengdu, China, 610000

[2.] Department of Chemistry, University of Cincinnati, Cincinnati, OH, USA, 45220

**Abstract**

Spermatogenesis is sustained by the proliferation and differentiation of spermatogonial stem cells (SSCs). SSCs are derived from primordial germ cells (PGCs) during embryonic development. The signal from the mammalian embryonic tissue is critical to the localization of PGCs and various factors and signaling pathways are involved into the regulation of these process. However, the molecules controlling these processes still need further clarification. In this concise review, the origin, development, differentiation of SSCs and regulating mechanisms throughout these processes are summarized and discussed in details. Finally, we mainly focused on elaborating the functions of TGF-β superfamily proteins and BMP4 in regulating SSCs formation and differentiation.

## Introduction

"Stem cell" was first used in 1901 by Regaud to describe the self-renewal ability of spermatogonia in testis (1). Regaud found that there is a stem cell system that constantly supplements differentiated cells. Today, it has been recognized that there are two types of stem cells: embryonic stem cells and adult stem cells. Adult stem cells are defined as cells that can form new progeny stem cells or more highly differentiated cells in a particular tissue. Many years ago, it was widely believed that there were four classic adult stem cell systems: skin, testis, small intestine, and bone marrow stem cells. The common feature of these tissues is that their cells are constantly updated to cope with the constant loss of cells. In recent years, the presence of adult stem cells has been found in more tissues, but its biological characteristics are likely to differ from the four classic adult stem cells described above(1).

Spermatogonial stem cells (SSCs) are a group of cells with high self-proliferative capacity and differentiation potential. It is the only stem cell in the male body that can transmit genetic information (2). The continuation of spermatogonial stem cells is the basis for the transfer of genetic material from generation to generation. In 1994, Brinster et al. developed a SSC transplantation experiment in mice; thereafter the study of SSCs has drawn a lot of attention (3). SSC transplantation is to transfer the normal animal or transgenic, mutant animal spermatogenic cells into the recipient testicular seminiferous tubules. The donor cells in the SSCs in the recipient testicles survive, migrate, settle and proliferate, thus start sperm generation, and even produce a technique of fertilization of sperm and progeny individuals.

The TGF-β superfamily protein consists of more than 40 members. Since each subunit has seven conserved cysteines, three disulfide bonds are formed inside the molecule, and one disulfide bond is formed between the subunits, thus during the signal transmission it forms a characteristic dimer, and target cells through the Ser/Thr kinase receptor binding function (4). The physiological processes it involved in the regulation include cell cycle progression, cell differentiation, reproductive function, development, exercise, adhesion, nerve growth, bone formation, wound healing, and immune monitoring and so on. The evolution of TGF-β family members among species is highly conserved which highlight the importance of TGF-β superfamily members.

## 1. Spermatogonial stem cells

In the testes, the term "stem cells" was first used by Regaud in 1901 to describe the self-renewing spermatogonia in the testes (1). Regaud recognizes that there must be a stem cell system in the testes that replenishes cells that are constantly differentiated. However, the term "stem cells" may have been used in the field of embryology as early as 1868(5). By definition, stem cells are primitive cells which self-renew and have the ability to differentiate into one or more other cell types (6). Nowadays, it is clear that there are two stem cells: embryonic stem cells and adult stem cells (1). Adult stem cells are self-renewing in specific tissues and organs, while at the same time forming cells with higher differentiation (1, 7). Many years ago, it was generally believed that there were four classic adult stem cells: skin stem cells, testicular stem cells, small intestinal stem cells and bone marrow stem cells. The common features of these tissues are: cell self-renewal in response to persistent cell loss. For example, spermatozoa may only be replaced by new sperm cells produced by sustained proliferation and differentiation of stem cell systems (7). In recent years, scientists have found stem cells in many tissues of the body, but their biological characteristics may be different from the above four types of classic stem cells (7). Human embryonic stem cells are derived from the inner cell mass of the blastocyst and are versatile, and they have the ability to differentiate into all cell types, but pluripotent stem cells usually do not form embryonic tissues such as amnion, chorionic, and other parts of the placenta. Competent stem cells have the ability to develop into more than one cell type, but are confined to a particular tissue, organ and physiological system. A typical example of pluripotent stem cells is hematopoietic stem cells that can differentiate into various types of blood cells, but cannot develop into brain cells or other types of cells; single-stem cells mainly refers to only along a single development path to form a specific cell type of stem cells (7). The focus of research on testicular stem cells has recently focused on its ability to reverse the formation of pluripotent stem cells; early, it is generally believed that testicular stem cells are monogensible stem cells, but this view received serious challenges recently. There is now strong evidence that in vitro culture of testicular stem cells (spermatogonial stem cells) can reversely differentiate into pluripotent stem cells that can become all cell types that form three germ layers without any genetic level operation (8-10).

1.1. Pluripotency of adult somatic cells and stem cells

Over the past few years, several research groups have published high-quality essays to elucidate that adult skin cells and other types of adult cells can be reprogrammed to form ES-like cells that are called induced pluripotent cells (Induced pluripotent stem cells, iPS) (7). Most of these studies were performed by using lentiviral vectors carrying important stem cell genes such as Oct4 (Pou5f1), Sox2, c-Myc and Klf4 (11-17). And by detecting Nanog (14, 16) or Oct4 (16) gene expression, or by observing the formation of ES-like clones by morphological methods, iPS cells can be screened (13). Although these cells are few in number, they can differentiate into germ cells, and epigenetically, they do not differ from the ES cells derived from the cell mass within the embryo (7). Recently, studies with mouse models have been demonstrated to modify the iPS cells induced by autologous skin cells at the gene level, and then to treat the sickle-type anemia mouse model by induction of differentiation (12). However, despite advances in the treatment of stem cells, which have attracted a wide range of interest, the instability of the stem cells and the retrovirus-infected stem cells cannot be used in the clinical treatment of humans, bringing the prospect of this technology to uncertain, which stimulate researchers to actively explore more close to the physiological means. In fact, scientists have recently been able to produce iPS cells by using only two transcription factors OCT4 and C-Myc or Klf4 to induce neural stem cells (18). More strikingly, a recent report shows that fibroblasts or hepatocytes can be directly induced to produce iPS cells by lentivirus infection (19). In addition, experimental animals, including primates, can obtain pluripotent embryonic stem cell lines by means of parthenogenesis, but this research is still in its early stage (20-22).

1.2 Spermatogonial stem cells

Spermatozoa occur in mammals starting in adolescence (in rodents and humans, respectively 5-7 days after birth and 10-13 years of age) and continue in the entire life (7). The process of sperm production is a complex process of cell differentiation, in which various types of cells in the testis are widely involved. This process is initiated and maintained by a small group of spermatogonial stem cells (SSCs) located in the basement membrane of the seminiferous tubules (23). SSCs are derived from primordial germ cells (PGCs) during embryonic development. The signal from the mammalian embryonic tissue is critical to the localization of PGCs (24). The mouse PGCs first appeared on days 7.25-7.5 of the embryonic development, which was located in the embryonic outer mesoderm (24) at the bottom of the original end of the urinary bladder, which has a high level of membrane alkaline phosphatase (ALP) activity (25). The mouse embryos after implantation confirmed that the ectodermules of 1-2 cells in diameter adjacent to the embryonic exosomes were the true birthplace of PGCs (26). PGCs continue to proliferate and form once every 16 hours until colonization of the gonadal (24). Female PGCs entered the first meiosis at E13.5 days and stagnated at the first meiotic prophase. Male PGCs stopped proliferating at E14.5 days until 1-2 days after birth to re-enter the proliferation cycle. The PGC of the proliferative resting stage is called gonocyte, which is located in the lumen of the seminiferous tubules(24). 1-2 days after birth, the gonocytes re-enter the spermatogenesis, and within 6 days after birth they all moved to the basement membrane, and this time the germ cells are known as SSCs. Based on the biological characteristics of spermatogonial cells with cytoplasmic transparency, high nuclear specificity and localization in the basement membrane, it has been possible to identify spermatogonia by morphological methods from the complex cellular components of seminiferous tubules (27). Early studies (27, 28) have shown that there are two types of stem cells: A0 (supplemental stem cells) and A1-A4 (renewed stem cells) in the testes of non-primates. Later, another model was proposed in which Asingle (AS) was thought to be a stem cell during spermatogenesis (29, 30). AS spermatogonia is present in a single presence, which can split into new AS stem cells or form a differentiated Aparied (Apr) [40-41connected by intercellular bridging; Apr further cleaves 8-, 16- or 32 - cells of Aal spermatogonia and further differentiated to form A1-A4, intermediate, B-type spermatogonia and primary spermatocytes (7). Although the differentiation of Apr and Aal spermatogonia is significantly higher than that of As, it is still considered that As, Apr and Aal are undifferentiated spermatogonia(7).

## 1.3 SSC surface markers

The research on molecular markers of human spermatogonia is progressing slowly. Recently, a report shows that monkeys and rodents are very similar in spermatogonocyte markers; these markers include PLZF, VASA, DAZL and GFRa1 (31); and reports on molecular markers of rodent SSCs have been reported (32-38). POU5, also known as OCT3/4, is thought to be a marker molecule for SSCs and its precursor nuclei (34, 39-41), and more importantly, it is possible that SSCs are more necessary for transcription factors (42). In addition, PLZF has been shown to be a nuclear transcription factor of SSCs (43, 44); He et al. Have demonstrated by immunostaining that OCT4 and GFRa1 co-located in the same subclass of mouse spermatogonia (7, 45). In addition, GPR125 is a marker molecule of mouse SSCs or their precursor cells (10); recent studies by Dym M et al. show that GPR125 is also a marker molecule for human SSC (7).

## 1.4 Regulation of the number of spermatogonia

The number of spermatogonia cells is regulated by the support cells (Sertoli cells) (46). Because the size of the testes is determined by the number of Sertoli cells; it is speculated that each Sertoli cell supports a certain amount of germ cells and spermatogonia (46). Meachem S et al. suggested that Sertoli cells constitute the microenvironment (niches) that support the SSCs survival, and these niches determine the presence or differentiation of a certain number of SSCs at specific times (46). In 2000, Xie and Spradling reported that stem cells were planted in this niche by studying Drosophila ovarian somatic cells (47). In addition, the proliferation pattern of spermatogonia after transplantation also supports the presence of germ cell niche (48, 49). Stem cell colonization and occupying niche's behavior may be spontaneous or dependent on certain genetic procedures. Transplanting the germ cells of dogs, pigs, rabbits and large animals into the seminiferous tubules of mice can be successfully colonized, but cannot complete differentiation (50, 51). The proliferation and differentiation of spermatogonia are at least partially regulated by hormone signaling (7). A large number of studies using different hormones models *in vivo* and *in vitro* and three-dimensional technology to quantify the spermatogonia to determine the different hormones on the promotion or inhibition of spermatogonia (1). The appearance of objective and fair counting methods paves the way for estimating the number of spermatogonia (52). The FACS method has been used to assess the proliferation and survival of spermatogonia (1). The re-colonization of germ cells into the seminiferous tubules by transplanting provides a very sophisticated approach to the development of testicular microenvironment regulating spermatogonia (1).

It has been clearly shown that FSH plays a major role in the regulation of spermatogonia through germ cell recovery and maintenance trials in rat testes (53-60). FSH may play a role in promoting the survival and division of spermatogonia (7). There is still a need for further study on the precise role of FSH in the development of spermatogonia. Experimental results of selective inhibition of FSH in the testis of adult rats by passive immunoassay showed a time-dependent decrease in early germ cells. This result suggests that A3-A4 spermatogonia (corresponding to seminiferous tubules

XIV-I) is the primary site of action for FSH (7). This stage corresponds to the highest level of FSH receptor (61) and FSH receptor mRNA expression (62). FSH-induced Sertoli cell cAMP reactivity also reached its peak at this stage (63). In primates, removal of Gonadotropin (GnRH antagonist treatment for 16-25 days) and then subjected to male hormones for contraception (20 weeks), the number of B-type spermatogonia cells was reduced to 10% of the control group (64, 65). These results showed that spermatogonia are the main role of gonadotropin to stimulate the target. Whether the inhibitory effect on spermatogonia is associated with the reduction of FSH and testosterone is still unknown; however, trials with rats, capes (66, 67) and rhesus (68) have confirmed that this role is indeed due to the elimination of FSH. The specific locus of spermatogenesis in humans and monkeys is still not very clear. Some studies have shown that withdrawal of FSH (66, 69) and gonadotropin (70) affects type A spermatogonia of monkeys. According to Schlatt and Weinbauer (1994), injection of GnRH antagonists significantly inhibited the proliferation of spermatogonia (PCNA), suggesting that the first step of mitotic (Apale spermatogonia stage) is the site of hormone impact. In contrast, Marshall et al. (1995) speculate that FSH treatment can selectively amplify type B rather than A pale spermatogonia in the removal of pituitary testosterone instead of rhesus monkeys. However, there is no clear evidence in the rat that FSH acts directly on type B or pre-fine spermatocytes (57).

The role of testosterone in spermatogonial development is unclear. There is no evidence that testosterone can support the development of spermatogonia in gonadotropin-treated rats (56, 71). Meistrich et al (1994) showed strong levels of testosterone concentrations in the testis that are detrimental to the development of spermatogonia and inhibit testosterone in irradiated rat and spermatogonia-induced adolescent mutant mouse models is necessary to promote spermatogonia development (72). Consistent with this study, treatment with GnHR antagonists in transplant experiments improves the colonization efficiency of donor cells in the recipient testis and the transient reduction in testosterone concentration can achieve this effect(73).

Stem cell factor (SCF) and its receptor c-Kit play an important role in the development of spermatogonia. Mutations in the SCF or c-Kit gene can lead to infertility due to migration, proliferation and survival defects of PGC [17]; destruction of SCF function prevents mitosis of spermatogonia. The injection of C-Kit antibodies into the testes of adult mice can lead to the loss of all differentiated spermatogonia. This result suggests that the differentiated spermatogonia is dependent on the C-Kit, whereas the undifferentiated cells are independent of the C-Kit (74). The important role of the SCF-C-Kit system has been demonstrated by transplanting spermatogonia from a sterile mouse carrying a mutated SCF gene into testis carrying a mutated C-Kit gene (7).

1.5 Identification of spermatogonial stem cells

In addition to identifying SSCs by labeling molecules, Brinster et al. established a mouse spermatogonial stem cell transplantation technique in the mid-1990s to identify the viability and function of SSCs cultured in vivo or in vitro (3, 75-77); Brinster and his colleagues also demonstrated that proliferation and spermatogenesis can also be accomplished by transplanting SSCs from other species into the testis of nude mice (78, 79). Nagano et al. demonstrated in 2002 that human spermatogonia (possibly SSCs) was transplanted into the testis of nude mice and colonized and cloned (80); recently, this method was extended to monkeys (31), it is assumed that germ cells of monkey SSCs are transplanted into nude mice treated with sodium nitrite and these cells can be colonized. The monkeys and human germ stem cells are transplanted into the testis of nude mice and can be colonized in the testis of the mice and cloned, but the process of spermatogenesis stops in the stage of spermatogonia (50, 51, 78, 80, 81).

1.6 SSC differentiation and dedifferentiation potential

As noted above, spermatogonias in rat and mouse testes are generally considered to be stem cell subclasses (82); however, Clermont *et al.* proposed another model of spermatogonial cell self-renewal (27, 28, 83, 84). In this model, a higher degree of differentiation of spermatogonia, especially A4-type spermatogonia can be restored into stem cells. Initially, this proposal was not widely praised by scientists in the field of spermatogonial research because it was difficult to accept a differentiated cell, especially A4 stage spermatogonia to differentiate into stem cells. Later in 2004, Brawley and Matunis published a book entitled "Regeneration of male germ line stem cells by spermatogonial dedifferentiation in vivo", which supported Clermont et al. In addition to a group of stem cells (real stem cells) that actually perform a function, the Yoshida group reported another spermatogonia group with potential self-renewal (85); the authors named the cells "potential stem cell populations ". These cells are highly differentiated spermatogonia or belong to Apaired, Aaligned or A1-A4 spermatogonia. The ability of dedifferentiation of differentiated spermatogonia seems to prove findings of Japanese (8) and Germany (9) scientists that spermatogonia (SSCs or their precursor cells) in mice can be biochemically induced and reprogrammed to form ES-like cells. This finding, confirmed by a US research group, found that SSCs and their precursors did reverse the formation of pluripotent ES-like cells (10). Golestaneh et al. have recently confirmed that there is a similar phenomenon in human testes (86). Thus, human SSCs have great potential to be used to induce ES-like cells and could be ultimately used in the treatment of human diseases.

1.7 Potential application of spermatogonial stem cells

Pluripotent stem cells (such as ES cells) have the potential to differentiate into various tissues and cells of the body and are therefore likely to be the source of organ tissue for transplantation. However, the practice of obtaining ES cells by artificial insemination inevitably leads to a series of ethical issues. The production of pluripotent stem cells (iPS cells) through the patient's own somatic cells is undoubtedly the best solution to this problem. There was no difference in morphology, proliferation and pluripotency between iPS cells and ES cells by teratoma formation and chimeric formation experiments. If iPS cells can be produced by somatic cells, the use of these iPS cells will inevitably lead to significant advances in new drug screening and regenerative medicine. SSCs can spontaneously reverse the differentiation (without the need to add foreign genes or reverse transcription virus as a carrier into the foreign gene) to form pluripotent stem cells. Thus, SSCs-derived ES-like cells can serve as a safer pathway for pluripotent stem cell sources than other adult cells. This highlights the potential value of human SSCs in cell-based autologous organ regeneration and treatment of various diseases.

**2. TGF-β superfamily members and their regulation of embryonic testicular development**

2.1 Overview of classic TGF-β superfamily signaling pathways

The TGF-β superfamily protein is composed of more than 40 members (4). TGF-β superfamily members are structurally related but different in function. Their physiological processes involved in regulation are cell cycle, cell differentiation, reproductive function, development, exercise, adhesion, nerve growth, bone formation, wound healing and immunization Monitoring, etc. (87-93). The evolution of family members among species is highly conserved to highlight the importance of TGF-β superfamily members. From the nematode to the human, they are found in the TGF-β superfamily homologues, their membrane receptors and involved in the transmission of the common receptor is also highly conservative. The mammalian TGF- β superfamily can be further divided into three subfamilies: TGF-β, activin/inhibin/nodal and BMP (94-97). Each subfamily consists of some functionally similar but non-overlapping isomeric members. TGF-β superfamily ligands are composed of homologous and heterodimers and their structures contain seven ordered cysteine residues. Six cysteine residues form a subunit disulfide bond that is critical to the integrity of the structure, and the remaining one cysteine residue is involved in the formation of disulfide bonds between subunits that maintain a stable dimer interface. TGF-β superfamily ligands are expressed in tissue-specific patterns and act through endocrine, paracrine, or autocrine manners. The specificity of the receptor, tissue distribution and expression levels all affect the final outcome of cellular responses.

The biological function of the cells for most of the TGF-β ligands is mediated by two single transmembrane Ser/Thr receptors, both of which are referred to as Class I and Class II receptors (98). Class I and Class II receptors are about 55 kDa and 70 kDa glycoproteins, which mediate signaling through binding to ligands. The Class I receptor is common to have a highly conserved TTSGSGSG cytoplasmic domain known as the GS region, which has the function of modulating the activity of class I receptor kinase. The first identified TGF-β family of proteins is the type II activin receptor (ActRII) (99), and a large class of Ser/Thr receptors has been identified subsequently, and their structural features are similar with activin receptors. To date, four Class II receptors in mammals have been identified: ActRIIB (100), AMHR-II (101, 102), TβRII (103) and BMPRII (104-107). In addition, seven Class I receptors (108-113) have been cloned and named as activin receptor-like kinase 1-7 (ALK1-7). Class I and Class II receptors differ in affinity and ligand specificity, and they may be differentially expressed to activate different types of intracellular pathways.

A TGF-β ligand binds to Class I and II membrane receptor and brings the two together to form a total trimer complex to initiate intracellular signaling (114, 115). The difference in assembly kinetics may lead to differences in the total complex of the ligand-receptor formation. Some members of the BMP family (the largest subfamily of the TGF-β ligand) can bind to class I or class II receptors to promote the formation of complexes; other TGF- β family members, including TGF- β and Activin must be combined with class II receptors to recruit class I receptors. Once the complex is formed, the Class II receptor is continuously activated by the kinase activity of the phosphorylated class I receptor in the GS domain, the class I receptor then phosphorylates the cells within the Smads signaling protein. There are 8 different Smad proteins, which can be divided into three categories: receptor Smad (R-Smad), co-modulate Smad (co-Smad) and inhibitory Smad (I-Smad) (116). R-Smad (Smad1, Smad5, Smad8 for BMP; Smad3, Smad2 for other TGF-β ligands) is directly phosphorylated and activated by the kinase activity of class I receptors, and the activated R-Smads form homotrimers and Co-Smad, Smad4, to form a heterotetramer. The activated Smad complex enters the nucleus and other nuclear factors to regulate the transcription of the target gene. I-Smad (Smad6, Smad7) negatively regulates TGF- β signaling by binding to R-Smad to bind the receptor or Co-Smad and direct the degradation of the target protein.

2.2 BMP family members on the embryonic germ cell development mediation

As mentioned earlier, gonads and germ cells in the early stages of embryonic development of mammalian embryos are dual (male and female) potentials that have the potential for development of testes with spermatogonia or ovaries with oocytes. These cells that form gametes in the future are called primordial germ cells (PGCs). Mouse PGCs were first seen at 7.25-7.5 dpc (24, 117), characterized by typical round and surface ALP activity positive. PGCs are located at the proximal epiblast of the mouse embryos and are induced two days earlier (5.25-5.5 dpc) by the extracellular signal. As described below, this process is significantly dependent on the BMPs signal (118-120).

BMPs can be divided into multiple groups based on the similarity or homology of mature carboxyl terminal sequence sequences (24, 89). BMP4 and BMP8b that belongs to the DPP 60A subgroups, respectively, are expressed in embryonic exosomes and gastrula of mouse embryos before.5-7.5 dpc of the (24, 120). The disrupted expression of either of them can lead to the inability of PGCs to form (119, 121) and developmental deficits in allantois. Both of these structures originate from the common precursors in the proximal epinement of the gastrula (119). BMP2 is a close relative of BMP4, which is predominantly expressed in visceral endoderm at early (5.0-7.5dpc) embryos, and its specific inactivation can also lead to a significant reduction in the number of PGCs (24).

Analysis of various BMPs -/+ heterozygous strains has elucidated the interrelationships between their functions. The number of PGCs in BMP2 and BMP4 double heterozygous mice (BMP2 -/+, BMP4 -/+) was significantly less than that of the single heterozygous mice, indicating that the function of these two proteins had an overlapping effect [145]. BMP4 -/+ and BMP8 -/+ double heterozygous mice were not significantly different from BMP4 -/+ single heterozygous; and the PGC numbers in BMP2-/+ and BMP8 -/+ mice embryos was not significantly different form that in the single heterozygous of each of them. The above results showed that BMP8 and BMP4 or BMP2 protein function without overlapping effects. Although the DPP family and the 60A family members transmit signals significantly different in the PGCs formation process (121, 122), but all of them are able to influence this process.

In the process of induction of PGCs in epiblast cells, it is not clear through which signal complexes the BMP2, BMP4 and BMP8 transduce their signal. A comprehensive study by Sousa Lopes et al. (120)demonstrated that BMP4 produced by embryonic exoscogular layers induced the formation of PGCs in the ectoderm through the ALK2 receptor of the visceral endoderm. Interestingly, ALK2-deficient embryos did not form PGCs, and their number was also significantly reduced in ALK2 -/+ animals. This phenotype is similar to the previously mentioned BMP4 knockout animal, which can be restored by expressing constitutively activated ALK2 in the visceral endoderm (120). Specific inactivation of the ALK4 gene in mouse embryos can lead to disordered tissue of the ectodermal and embryonic exosomes, leading to abnormal development of the anterior oocysts (123).

Smads are also important for embryonic development. Smad1 expression is located in the ectodermal and visceral endoderm during the formation of the gastrula, but its immediate expression in 7.5 dpc wild type mice suggests that Smad1 may play a role in PGCs formation rather than proliferation or maintenance (124). Smad1 knockout mice died at 10.5-11.5 dpc (125), and these individuals produced only a small amount of PGCs, demonstrating that Smad1 is critical to the formation of PGCs (124, 125).

Smad5 knockout mice also died at 9.5-11.5 dpc (126) of the embryonic stage; similar to that of Smad1 knockout mice, they showed minimal or missing PGCs phenotype. Smad5 probably plays an important in determining which ectodermal cells are differentiated into PGCs (126, 127), because it can be detected in the normal mouse ectoderm at 6.5 dpc and in all three layers of the 7.5 dpc embryo. Smad1 and Smad5 were expressed in the ectoderm at the initial stage of gastrulation, but Smad8 was detectable only in the allantoic sac (125-128). Smad2 gene was widely expressed in mouse embryonic development (129-131).

Although Smad2 knockout mice died during the embryonic period of 7.5-8.5 dpc days, most of the individual's PGCs were still developed (125). This suggests that activation of upstream signals of Smad2, such as activins, TGF- β or nodal, may not be necessary for PGCs development. Consistently, Smad4, which is a vital protein involved in multifunctional critical signaling pathways, also widely expressed. Smad4 knockout resulted in visceral endodermal developmental defects and died at 6.5-8.5 dpc (132, 133). Smad4 knockout mice are similar to the phenotype of most other BMPs knockout mice (132), thus highlighting the role of Smad4 in the presence of numerous interacting TGF-β signaling pathways.

**3. TGF-β superfamily members participate in the spermatogenesis process after birth**

3.1 Cell differentiation during after birth spermatogenesis

Postnatal individual spermatogenesis is defined as the process of SSCs that develop mature sperm in the testicular seminiferous tubules. The spermatogenesis of the mice begins within one day of birth, at which time the reproductive cells are mitotic and migrate from the center of the seminiferous tubules to the peripheral basal membrane (134). These cells formed the first trimester cells on the 6$^{th}$ day after birth, and the germ cells that did not migrate to the basement membrane were subsequently apoptotic. In the course of a series of mitotic processes, the cytoplasm of spermatogonia is not completely split to form Ap, Aal, A1-A4 sperm and intermediate spermatozoa (In), and finally on the eighth day after birth, becomes permatogonia (Reviewed in (135)). The last mitosis of the B-type spermatogonia was the formation of the earliest meiosis cells, i.e., the pre-fine line spermatocytes (about 10 dpp in the mouse testes), which were isolated from the basement membrane and passed through Sertoli cells formed tight connection.

Sertoli cells are specialized "nourish" cells that provide nutritional and structural support for adult germ cell development. The proliferation of Sertoli cells lasted about 14 dpp, and then subjected to significant morphological and functional translations to reach maturation. The tight junction between the Sertoli cells formed at the earliest 7-14 dpp will isolate meiotic and post- meiotic germ cells from the factor in lymph and blood circulation as well as the secretions from the top surface of the seminiferous tubules. Thus, providing a special environment for the development of post-meiotic germ cells and discharge them from the testes (see reviewed in(4)).

The midpoint of meiosis begins in each spermatocyte division to form two diploid secondary spermatocytes, and then rapidly undergoes a second meiosis to form 4 haploid sperm cells, these cells interconnected by persistent cytoplasmic bridging bridges. The round sperm cells appear about 20 dpp for the first time, and undergo further differentiation and morphological changes, resulting in the formation of mature spermatozoa for the first time at about 35 dpp. This process is repeated in adult mouse testes. Thus, the A-type spermatogonia at each basement undergoes two mitotic meiosis and more than two weeks of differentiation, eventually forming spermatozoa and releasing the lumen into the center of the seminiferous tubules (reviewed in(82, 136)).

3.2 Spermatogenesis

The study of cultured organs and cells has deepened on understanding of the function of TGF-β superfamily members during spermatogenesis, especially in the differentiation of spermatogonia at initial stage of the first spermatogenesis process. The following is a brief overview of the functions of BMPs and GDNF in this process.

BMPs

There is evidence that Sertoli cells in adolescent rodents in the proliferative phase can produce or secrete activin, inhibin and BMP4, which regulate the activity of germ cells through co-receptors. The testes of early rodents were born with receptor subunits such as activin IIB, ALK2, ALK3 and ALK4 (137-140).

BMP4 was added to the 4-day-old mouse testis (which contained only undifferentiated spermatozoa) in an enriched cultured germ cell culture system that significantly induced upregulation of the functional C-kit receptor, resulting in the elevation of SSCs Differentiation. In another experiment, the addition of BMP2 and FSH to the 7-day-old mouse testicular fragment culture system could cause proliferation of spermatogonia (138).

TGF-β ligands can also affect the survival of germ cells before and during the initiation of spermatogenesis. TGF-β2 ligand and TGF-β receptor are expressed in neonatal rodent testicular germ cells (141). TGF-β1 or TGF-β2 can increase the apoptosis rate of germ cells in the testicular fragments of 3-day-old rats (142). TGF-β3 protein can also be detected in reproductive cells (143), but no reports have been published on the study of its functional in such cells. Downregulation of mRNA and protein levels of TGF-β3 appears to be necessary to establish a tight junction between Sertoli cells (144). TGF-β3 was added to the 20-day dpc rat Sertoli cell culture to disrupt its ability to form a tight junction, and the level of P38 MAPK phosphorylation was elevated, but the Smad protein level was not affected. In contrast, inhibition of phosphorylation of P38 MAPK in these cell cultures led to the formation of tight junctions, suggesting that TGF-β3 is involved in regulating the dynamic balance of tight junctions of Sertoli cells (144).

GDNF

The pro-inflammatory function of GDNF on undifferentiated reproductive stem cells has been demonstrated by transgenic/knockout and *in vitro* culture experiments (32). *In vitro* culture of germline stem cells derived from neonatal mouse testes were maintained only in the presence of GDNF(145, 146). The GDNF in the testes is produced and secreted by Sertoli cells, acting on the undifferentiated SSCs by the C-RET /GFRa1 receptor (147).

3.3 TGF-β superfamily members involved in the occurrence of sowing in adult animals

Researchers have established a number of different gene-modified animal models to detect the effect of TGF-β superfamily members on normal physiological functions (87). The phenotype of reproductive capacity has been identified in a variety of situations, which allows us to study the function of the TGF-β superfamily as a member at specific stages of spermatogenesis (87, 148-152). A few examples are summarized below:

BMPs

BMP7, 8a and 8b are expressed in the sperm and early spermatocytes of young testes and are also expressed in adult testicular round spermatozoa. Gene-specific inactivation tests showed that BMP8b was important for the initiation of spermatogenesis in young testes, whereas BMP8a did not affect the initiation of spermatogenesis, but it was important for the maintenance of spermatogenesis in adult testes. Knockout of BMP8a resulted in an increased apoptotic germ cell proporation at the meiotic stage in 20-50% of seminiferous tubules, a decrease in sperm production capacity and eventually lead to infertility. In BMP8a -/- and BMP7 -/+ heterozygous mice, this apoptitic meiotic germ cells were detected in 40-80% of the seminiferous tubules, indicating that BMP7 plays an adjuvant role in the maintenance of spermatozoa (153-157). BMP4 knockout mice died at birth and their testes were absent of germ cells (119); the heterozygous mice survived adulthood and had normal testicular phenotype at 4 weeks of age. However, adult heterozygous mice had a smaller testis, and the meiotic division of germline cells resulted in a decrease in the number and activity of spermatozoa, similar to the phenotype of BMP8a knockout mice (158). The knockout animal model also revealed the important role of BMP4, 7, 8a and 8b in maintaining epididymal function; the reduction in sperm motility in gene knockout mice may be due in part to epidural degradation.

GDNF

Genetically modified mice provide useful information for studying the function of GDNF in SSCs. In the testes, GDNF is secreted by Sertoli cells and regulated by FSH(147). Mice carrying a GDNF allele were depleted of the stem cell, indicating the maintenance of SSCs *in vivo* required a certain level of GDNF factors (32). These mice were able to initiate the first wave of spermatogenesis, but the testicular atrophy phenotype was observed at week 5, although many seminiferous tubules contained sperm cells, spermatogonia was absent.

In contrast, transgenic mice that specifically overexpress GDNF in testes are also infertile, which exhibit undiagnosed spermatogenic cell phenotypes, spermatogenic differentiation and eventually lead to germ cell apoptosis (32, 159-162). SSCs in this mouse model can proliferate, but they do not respond to normal differentiation-inducing signals (eg, retinoic acid, etc.), leading to the destruction of seminiferous tubules and failure of spermatogenesis. Untill 2-3 weeks, type A spermatogonias formed a large number of clusters in the testes. The ratio of testes/body weight of these mice were relative lighter at the age of 4-8 weeks compared to that of normal mice, and did not have reproductive capacity until 8 weeks. These germ cell populations were degraded at 10 weeks of age, leading to the presence of seminiferous tubules by only one layer of spermatogonia and no detectable presence of sperm cells. These mice eventually form non-metastatic testicular tumors (32, 163). Interestingly, when hGDNF was overexpressed in Sertoli cells of 12-day-old mice by *in vivo* electroporation mathod, mice testes, spermatogonia could alos form accumulation after four weeks. After transplantaion into the testosterone-treated mice, these cells were able to colonize seminiferous tubules, form clones and complete spermatogenesis, demonstrating there SSC property (164).

Results from transplantation assays showed that germ cells that exposed to high levels of GDNF could maintain their differentiation ability. Spermatogonial cells from GDNF Sertoli cells overexpressing mice could complete the spermatogenesis process after transplanted into normal testes (164). Once these cells were removed from this "super-stimulating" environment, they can be properly differentiated. As a method of *in vivo* amplification of SSCs, this clearly has the potential to provide a large number of SSCs for experimental studies.